\documentclass[aps,prl,twocolumn,superscriptaddress]{revtex4}
\usepackage{amsmath}
\usepackage{amssymb}
\usepackage{graphicx}

\begin{document}

\title{Double quantum dot as a probe of nonequilibrium charge
fluctuations at the quantum point contact}

\author{A. L. Chudnovskiy}
\affiliation{1. Institut f\"ur Theoretische Physik, Universit\"at Hamburg,
Jungiusstr 9, D-20355 Hamburg, Germany}

\date{\today}

\begin{abstract} 
Absorption of energy quanta generated by quantum point contact results in the inelastic current through the double quantum dot placed nearby. In contrast to a single quantum dot, the inelastic current through the double quantum dot is sensitive to the energy dependence of the quantum point contact  transmission, which can explain the experimentally observed 
features.  We calculate the inelastic current as a function of  microscopic parameters of the circuit.  
\end{abstract}


\maketitle

Double quantum dot (DQD) has been recently proposed as a detector of  nonequilibrium noise generated by nearby mesoscopic devices \cite{Aguado00}. 
This idea was experimentally realized in measurements of the nonequilibrium noise spectrum  of a quantum point contact (QPC) detected by DQD \cite{Khrapai,Gustavsson}. The experiments provided a lot of interesting and in some respect puzzling results. 

 The noise detection is based on the generation of inelastic 
current through DQD assisted by absorption of energy quanta emitted by QPC. To implement the noise measurement, DQD is brought into the Coulomb blockade regime with the highest energy electron localized in one of its dots hereafter referred to as dot 1.  The exchange of the electron between the two quantum dots brings DQD into the excited state, the excitation energy $\Delta$ being fixed by the gate voltages. 
Absorbing an energy quantum, an electron tunnels from the low energy state in  quantum dot 1 to the excited state localized in quantum dot 2. The tunnel barrier between the quantum dots is tuned to be much higher than the barriers between the dots and the adjacent leads, so that after each interdot tunneling event the electron almost immediately escapes into the adjacent electron reservoir. Another electron occupies quantum dot 1, and the system returns to the ground state, the unit of charge having  been transferred through DQD \cite{Khrapai}. The generated current is therefore proportional to the noise power on the excitation frequency of DQD.  
The nonequilibrium noise is generated by QPC, which is brought in a strongly nonequilibrium transport regime by application of transport voltage. At the same time, the plunger voltage applied to QPC controls its transmission. 

Theoretical calculations of the generated inelastic current have been performed in Ref. \cite{Aguado00}, where it has been related  to the nonequilibrium noise power $S_{I}(\Delta/\hbar)$ generated by QPC at the frequency $\Delta/\hbar$. This noise power is given by the local current fluctuations in an arbitrary spatial point of QPC, 
\begin{equation}
S^{\mathrm{local}}_{I}(\omega)=\int_{-\infty}^{\infty}d\tau e^{i\omega\tau} \langle \delta I(x,\tau) 
\delta I(x,0)\rangle.
\label{S_local}
\end{equation}
 Based on the energy conservation law one concludes that increasing the QPC transport  voltage $V_{\mathrm{QPC}}$, the current through DQD will start at the point
$V^*_{\mathrm{QPC}}=\Delta/e$, when the quanta with energy $\Delta$ appear
in the nonequilibrium noise spectrum \cite{Blanter}. A puzzling feature
of the experimental measurements
is the independence of the  threshold voltage $V^*_{\mathrm{QPC}}$  of the DQD
excitation energy $\Delta$ for a finite range of energies, contrary to the expectations based on the energy conservation law \cite{Khrapai}.

\begin{figure}
\includegraphics[width=8cm,height=6cm,angle=0]{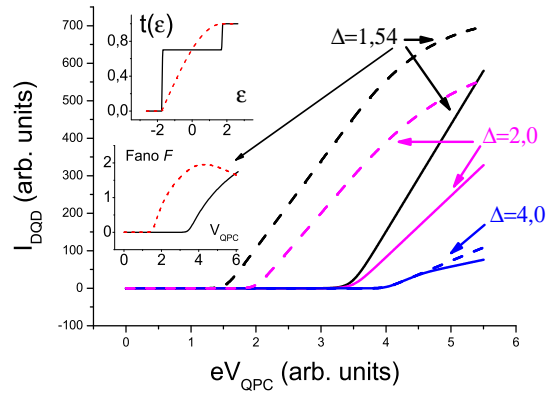}%
\caption{(Color online) Current through  DQD as a function of QPC bias voltage. 
Solid lines correspond to a plateau-like energy dependence of the QPC transmission
amplitude $t_{\epsilon}$, dashed lines correspond to a continuous
$t_{\epsilon}$ (see upper inset). 
For $\Delta$ smaller then the width of the plateau $E_0$ in $t_{\epsilon}$, 
the threshold voltage is defined not by $\Delta$ but by $E_0$ (the current
onset for the first two solid curves almost coincides).  For continuous
$t_{\epsilon}$ the current onset is determined by $eV_{\mathrm{QPC}}=\Delta$ 
(see the dashed curves).  Parameters of the interaction 
$U_{1}^f=U_{2}^{f}=1.0$. Other parameters are relevant for the
experiment in Ref. \cite{Khrapai}. 
Lower inset: dependence of the finite frequency Fano factor on $V_{\mathrm{QPC}}$. 
\label{figIVsd-delta}}
\end{figure}

In this Letter we provide a theoretical description of QPC--DQD system in the 
nonequilibrium regime that allows to relate experimental measurements of the
inelastic current to  microscopic parameters.  We show that since DQD  is an object extended over both sides of QPC, the local noise power  (\ref{S_local}) { \em is not the relevant quantity} for the inelastic DQD current. 
Rather, the noise power absorbed by DQD includes spatially nonlocal correlations of current fluctuations at positions of two quantum dots. The relevant voltage power is given by 
\begin{equation}
\mathcal{S}_{V}(\omega)=\left\langle \left\vert Z(x_1, \omega) \delta \hat{I}(x_1,\omega) 
+ Z(x_2, \omega) \delta \hat{I}(x_2,\omega)\right\vert^2 \right\rangle, 
\label{Vpower}
\end{equation}
where $Z(x_i, \omega)$  is the spatially dependent trans-impedance of the circuit,  $x_i$ denotes the position of the corresponding quantum dot, and 
 $\omega=\Delta/\hbar$ is the absorption frequency.     
 In the case of the symmetric coupling of both quantum dots to QPC, which is called a symmetric circuit in what follows,  the trans-impedance becomes independent of spatial coordinate. 
Then the inelastic current is given by the expression similar to that obtained in Ref. \cite{Aguado00} 
\begin{equation}
I_{\mathrm{DQD}}=\frac{e^3}{\hbar^2}\frac{\vert Z(\Delta/\hbar)\vert^2}{\Delta^2} 
\mathcal{S}_{I}\left(\frac{\Delta}{\hbar}\right),  
\label{Idqd-simple}
\end{equation} 
but with the {\em nonlocal} current noise power $\mathcal{S}_{I}(\omega)$ following from  
(\ref{Vpower}) 
\begin{equation}
\mathcal{S}_{I}(\omega)= \sum_{i,j=1}^2 \int_{-\infty}^{\infty}d\tau e^{i\omega\tau} \left\langle \delta \hat{I}(x_i,\tau) \delta \hat{I}(x_j, 0) \right\rangle.  
\label{S_I}
\end{equation} 
The  presence of nonlocal current correlations substantially modifies the resulting noise spectrum. So,  while the energy dependence of the QPC transmission is not essential for the local current fluctuations (\ref{S_local}) \cite{Blanter,Aguado00,Lesovik}, it becomes crucial for the spatially nonlocal fluctuations of the current  (\ref{S_I}). As the result, the generated inelastic current becomes sensitive to the energy dependence of QPC transmission. In particular, for the symmetric circuit the inelastic current turns to zero when the QPC transmission is energy independent within the transport voltage window. This can explain the independence of the DQD current onset voltage $V^*_{\mathrm{QPC}}$ on the DQD excitation energy $\Delta$ seen in experiment \cite{Khrapai}.  These findings are illustrated in Fig. \ref{figIVsd-delta}. If the QPC transmission amplitude exhibits a plateau in its energy dependence, the current onset voltage is determined by the width of the plateau as long as that width exceeds the DQD excitation energy $\Delta$ (solid lines). In contrast, the threshold voltage equals $\Delta/e$ for the continuously rising QPC transmission (dashed lines).

To emphasize the importance of the energy dependence of QPC transmission for the power absorbed by DQD, it is advantageous to define  an analogy of the Fano factor for a finite frequency noise 
$
\mathcal{F}(\omega)\equiv S_{I}(\omega)/(2eI_{\mathrm{QPC}})
$.  This value relates the power generated at frequency $\omega$ and defined by 
(\ref{S_I}) to the direct current through QPC.  Its explicit expression reads 
\begin{eqnarray}
\nonumber && 
\mathcal{F}(\omega)=\left[1+ 
\tanh\left(\frac{\hbar\omega}{2T}\right)\coth\left(\frac{eV
-\hbar\omega}{2 T}\right)\right]\times \\
&&
\frac{\int d\epsilon
\left\vert r_{\epsilon+\hbar\omega}t_{\epsilon}-t_{\epsilon+\hbar\omega} 
r_{\epsilon}\right\vert^2 \left(f^{L}_{\epsilon+\hbar\omega}- 
f^{R}_{\epsilon}\right)}{2 \int 
d\epsilon \vert t_{\epsilon}\vert^2\left(f^{L}_{\epsilon}-f^{R}_{\epsilon}\right)}, 
\label{F_omega}
\end{eqnarray}
where $t_{\epsilon}$ and $r_{\epsilon}$ are the transmission and reflection amplitudes of QPC at energy $\epsilon$. The Fermi distributions in the left (source)/right (drain) reservoirs of QPC  are denoted as $f^{L/R}_{\epsilon}$. Their chemical potentials differ by the QPC transport voltage and can be written as $\mu_{L/R}=\pm eV_{\mathrm{QPC}}/2$, the chemical potential in the unbiased QPC being taken as zero. 
For zero temperature, and for the absorption frequency $\omega=\Delta/\hbar$ 
Eq. (\ref{F_omega}) simplifies to 
\begin{equation}
\mathcal{F}(\Delta/\hbar)=\Theta(eV-\Delta)\frac{\int_{-\frac{eV}{2}}^{\frac{eV}{2}-\Delta} d\epsilon
\left\vert r_{\epsilon+\Delta}t_{\epsilon}-t_{\epsilon+\Delta} 
r_{\epsilon}\right\vert^2}  
{2 \int_{-eV/2}^{eV/2}d\epsilon \vert t_{\epsilon}\vert^2}. 
\label{F-simple}
\end{equation} 
It is evident from (\ref{F-simple}), that the Fano factor is determined by the energy dependence of QPC transmission.  
The dependence of the finite frequency Fano factor on the QPC transport voltage is shown in the inset to Fig. \ref{figIVsd-delta}.  The energy threshold of the Fano factor corresponds to the width of the plateau in QPC transmission. Furthermore, the Fano factor drops if the applied voltage is larger than the energy interval for the onset of the conducting channel. In that case, the DQD inelastic current reaches saturation while the direct QPC current continues to grow. 
Using the Fano factor (\ref{F_omega}), the expression for the generated inelastic current can be cast in the form 
\begin{equation}
I_{\mathrm{DQD}}=2e I_{\mathrm{QPC}} \mathcal{F}(\Delta/\hbar)
\frac{\vert Z(\Delta/\hbar)\vert}{\Delta^2} \Gamma_{\mathrm{DQD}}
, 
\label{I_DQD-Fano}
\end{equation}
where $\Gamma_{\mathrm{DQD}}=\frac{8e t_0^2}{\pi\hbar R_Q}$ is the DQD rectification factor, $t_0$ denotes the tunnel amplitude between the two quantum dots,  and  $R_Q=h/e^2$ is the quantum resistance. 

In what follows we introduce the theoretical model for the coupled QPC -- DQD system and outline the derivation of the presented results. Since the maximal inelastic current is observed when a new conducting channel is opening in QPC, we concentrate on a single conducting channel of QPC. 
We distinguish two species of electrons in QPC, namely those coming from the right and the left reservoirs, and  we describe them by the fermion field operators $\psi_{R/L} (x)$ \cite{indices}.  The electrons of each sort are in equilibrium with its own reservoir. Taking the position of  the QPC potential barrier at $x=0$, we represent the field on each side of it as 
\begin{eqnarray}
\nonumber &&
\psi(x)=\int\frac{dk}{2\pi}\left\{ \psi_L(k)\left[e^{i(p_F+k)x}+ 
r_{\epsilon} 
e^{-i(p_F+k)x}\right] \right. \\
&& \left. 
+\psi_R(k) t^*_{\epsilon}e^{-i(p_F+k)x}\right\}, \ \mbox{for} \ x<0,   
\label{psi_s} \\
\nonumber && 
\psi(x)=\int\frac{dk}{2\pi}\left\{
\psi_R(k)\left[e^{-i(p_F+k)x}+r^*_{\epsilon} e^{i(p_F+k)x}\right] 
\right. \\ 
&& \left. 
+\psi_L(k) t_{\epsilon} e^{i(p_F+k)x}\right\}, \ \mbox{for} \ x>0.   
\label{psi_d}
\end{eqnarray}
Here $p_F$ denotes the Fermi wave vector at zero transport voltage. 

The structure of the interaction between DQD and QPC-channel plays a crucial
role for the generation of inelastic current, determining the trans-impedance $Z(x_i,  \omega)$. Due to the presence of external electrodes, the effective interaction becomes screened and time-retarded. 
Moreover, the presence of QPC violates the spatial homogeneity of the interaction. 
Therefore,  we can write the interaction term in the action in the form 
\begin{eqnarray}
\nonumber &&
\mathcal{A}_{\mathrm{int}}= -e^2\sum_{i=1,2}\int d x \int dt dt' U_i(\vert x \vert, 
\vert x - x_i 
\vert, t-t') \times \\ 
&& 
\hat{n}(x,t) \hat{n}_i(t') 
\label{S_int}
\end{eqnarray}
Here $\hat{n}_1$ and $\hat{n}_2$ are the particle number operators in each 
quantum dot, and $\hat{n}(x)$ is the operator of density fluctuations in the conducting channel  at point $x$.  We assume that the quantum dots are situated far away from the potential barrier of QPC, one on each side of it (see experimental setup 
of Ref. \cite{Khrapai}).
In the detection regime, the total occupation of DQD is fixed to $n_1(t)+n_2(t)=1$. 
This allows us to use a pseudo-spin 1/2 description of DQD. We associate the states localized in the quantum dots 1 and 2 with the spin-up and spin-down states respectively. The charge transfer between the two quantum dots corresponds to the spin-flip between the ground state spin-up and the excited state spin-down. The interaction term 
(\ref{S_int}) can be splitted into the interaction with the total charge of DQD, and the interaction with the $z$-component of DQD
pseudo-spin, $\hat{S}^z=\hat{n}_1-\hat{n}_2$.   Note that the product 
$\hat{\pi}=e \hat{S}^z$,  is proportional to the operator of the DQD dipole moment. 
Omitting the interaction with the total charge of DQD, the relevant interaction can be represented as the one between the dipole moments of DQD and QPC, which in the Fourier-transformed form reads 
\begin{equation}
\mathcal{A}_{\rm int}=-\int\frac{d\omega}{2\pi}  \hat{P}(\omega) 
\hat{\pi}(-\omega),
\label{intdipole} 
\end{equation}
where the QPC dipole moment 
$\hat{P}(\omega) $ is defined as 
\begin{eqnarray}
\nonumber && 
\hat{P}(\omega) =e\int d x \hat{n}(x,\omega)\left[U_1\left(\vert x \vert, \vert x -
x_1\vert, \omega\right) \right. \\
&& \left.
-U_2\left(\vert x \vert, \vert x - x_2\vert, \omega\right)\right]. 
\label{defP}
\end{eqnarray} 
The forward and backward inelastic scattering amplitudes are the only relevant ones for the one-dimensional motion of electrons in QPC.  
They are given in terms of Fourier transforms  
$\int d x  e^{-iq( x- x_i)}U_{i}(\vert x\vert, \vert x- x_i\vert)$ at wave vectors 
$\vert q\vert \ll p_F$ and $q=\pm 2p_F$. We assume that the interaction is strongly screened, and it takes place only in a small region of the size of the screening  length around each quantum dot.  Then the behavior of scattering amplitudes at small wave vectors is smooth, and we can approximate $U_i(\vert q\vert \ll p_F)\approx U_i(q=0)\equiv U_{i}^{f}$ for the forward scattering.  
 For the backward scattering we obtain    
$U_{i}^b(\pm 2p_F)=U_{i}(\pm 2p_F)e^{\mp 2ip_F x_{i}}$ 
Taking into account the finite size of a quantum dot, one has to integrate over $x_i$ within that size, which greatly diminishes the back-scattering amplitude because of the rapidly oscillating factors 
$e^{\pm 2ip_F x_i}$.  On that account we neglect the back-scattering amplitudes in what follows. 

Using Eqs. (\ref{psi_s}), (\ref{psi_d}),  (\ref{defP}),  we
represent the total dipole moment operator $\hat{P}$ in terms of the fields of right- and left-reservoirs  as follows 
\cite{footnote3}  
\begin{equation}
\hat{P}(\omega)= e\sum_{\chi, \chi'=R,L} \int\frac{dk dk'}{(2\pi)^2} 
w^{\chi' \chi}_{\epsilon', \epsilon} \psi_{\chi'}^+(k') \psi_{\chi}(k)
\delta(\epsilon'-\epsilon-\omega). 
\label{int-chiral}
\end{equation}
The combined effect of the scattering by QPC potential barrier and
interactions with quantum dots is captured by the effective inelastic
scattering amplitudes $w^{\chi'\chi}_{\epsilon', \epsilon}$ between the two species of fermion fields. Further calculation shows that only the
inelastic scattering between different species contributes to the nonequilibrium noise power.  The amplitude $w^{LR}$ is given by 
\begin{equation}
w^{LR}_{\epsilon', \epsilon}= U_{1}^f r_{\epsilon'}t_{\epsilon} 
-U_{2}^f t_{\epsilon'}r_{\epsilon}, 
\label{wRL-2} 
\end{equation}     
and the amplitude $w^{RL}_{\epsilon', \epsilon}$
is   
obtained from $w^{LR}_{\epsilon', \epsilon}$ by exchange $U_{1}^f \leftrightarrow
-U_{2}^f$. 
Using the representation of the electric current operator in the basis (\ref{psi_s}), (\ref{psi_d}) we can express the dipole moment $\hat{P}$, in terms of the Fourier transform of the current operator at frequency $\omega$ \cite{Martin,Kamenev-Levchenko}, 
$
\hat{P}(\omega)=\sum_{i=1,2} Z(x_i, \omega) \hat{I}(x_i, \omega) 
$ 
introducing the spatially dependent trans-impedance  
$
Z(x_i,\omega)=\frac{1}{v_F}U_{i}^f  
$. 
At this point it becomes evident that the dipole moment interacting with DQD involves spatially nonlocal correlations of QPC current. 
In the case of symmetric circuit, $U_{1}^f=U_{2}^f$, the trans-impedance becomes independent of coordinate. Its expression  in terms of the elements of effective electric circuit is provided in Ref. \cite{Aguado00}. 

The generated inelastic current is calculated perturbatively in the lowest order of QPC-DQD interaction employing the  Keldysh technique \cite{Keldysh,Kamenev}. The total action is given by $\mathcal{A}=\mathcal{A}_0+\mathcal{A}_{\mathrm{int}}$, with the interaction part given by (\ref{intdipole}) and the free part 
\begin{equation}
\mathcal{A}_0=\int dt \left\{\sum_{\chi=R,L} \overline{\psi}_{\chi}G^{-1}_{\chi}\psi_{\chi} 
+ \overline{\Phi}{\bf D}^{-1}\Phi\right\}.
\label{S-Keldysh}
\end{equation} 
In the free part of the action,  we used the semifermionic representation of the pseudospin degrees of
freedom of DQD \cite{Popov-Fedotov,Kiselev}. Here $\Phi=(\phi_{+}, \phi_{-})^T$ is the spinor Grassmann field. Each spin component $\phi_{\sigma}$ is in turn a two-component  field in the retarded-advanced space.  ${\bf D}$ is the
semifermionic Green's function, its retarded components are given by 
$D^R_{\sigma\sigma}(\epsilon)=[\epsilon-\sigma\Delta/2+io]^{-1}$,
$D^R_{\sigma,-\sigma}=\frac{t_0}{\Delta} 
\left(D^R_{++}-D^R_{--}\right)$. The
semifermionic spin-1/2 representation imposes special rules of calculus
in Keldysh formalism.  The Keldysh component is parameterized as $D^K(\epsilon)=D^R(\epsilon)
F_s(\epsilon)-F_s(\epsilon)D^A(\epsilon)$  with the function
$F_s(\epsilon)=\tanh\left(\frac{\epsilon}{T}\right)\pm
\frac{i}{\cosh\left(\frac{\epsilon}{T}\right)}$. In the diagrammatic expansion, each diagram is calculated taking  $F_s$ once with the plus and once with the minus 
sign, and the half-sum of the results is taken at the end of the calculation \cite{Kiselev}. 

The Green's function of the fermions from the reservoir $\chi$ in QPC channel is
given by 
$G_{R/L}^R(\epsilon, k)=[\epsilon\mp v_F k -\mu_{R/L}+io]^{-1}$, the Keldysh component is
obtained by usual rules \cite{Keldysh,Kamenev}.  The energy $\epsilon$ is counted from the Fermi level of the unbiased QPC. 

\begin{figure} 
\includegraphics[width=7cm,height=5cm,angle=0]{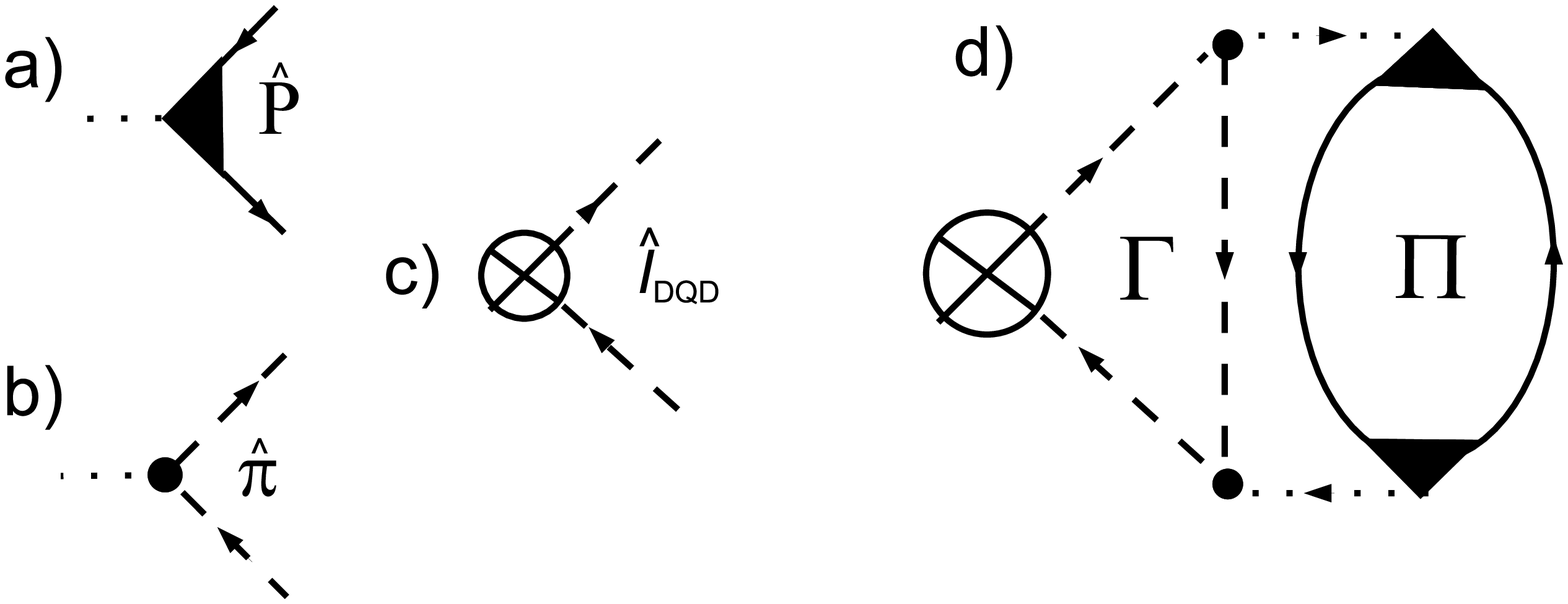}%
\vskip -2cm
\caption{Diagrammatic representation: 
a) Vertex of QPC dipol moment $\hat{P}$; b) Vertex of DQD dipol moment $\hat{\pi}$;  
c) DQD current vertex $\hat{I}_{\mathrm{DQD}}$; d) Diagram for DQD current in the second order of interactions. $\Pi$ denotes the QPC polarization operator, and $\Gamma$ is the DQD rectification factor.
Dashed lines denote the semifermionic Green's functions ${\bf D}(\epsilon)$; 
Solid lines denote the Green's functions in QPC. 
\label{fig-Idqd-diag}}
\end{figure}

The operator of the current through the double quantum dot can be written in the pseudospin representation as  
$
\hat{I}_{\mathrm{DQD}}=-e t_0\hat{S}^y/\hbar.
$ 
In the Keldysh formalism, its quantum component is proportional to the
matrix 
$\sigma^x$ acting in the retarded-advanced space $
\hat{I}_{\mathrm{DQD}}^{\mathrm{q}}=\sigma^x\otimes\hat{I}_{\mathrm{DQD}}
$.
The current is calculated using a diagrammatic representation derived from the
Keldysh path integral with the action  (\ref{S-Keldysh}) 
(see Fig. \ref{fig-Idqd-diag}).  The diagram 
for the current in the second order of perturbation in the
interaction is shown in Fig. \ref{fig-Idqd-diag}d), the corresponding analytical
expression reads 
$
I_{\mathrm{DQD}}=
Tr\left\{\hat{\Gamma}\hat{\Pi}\right\}.   
$
The polarization operator 
$\hat{\Pi}=\left\langle \hat{P}(\omega) 
\hat{P}(-\omega)\right\rangle$  can be represented in terms of current operators and trans-impedances as $\hat{\Pi}=\left\langle\vert\sum_{i=1,2}Z(x_i, \omega)\hat{I}(x_i, \omega)\vert^2\right\rangle$. For the symmetric circuit  
we obtain 
$
\hat{\Pi}(\omega)=\vert Z(\omega)\vert^2 \mathcal{S}_I(\omega) 
$ with the noise power $\mathcal{S}_I(\omega)$ given by (\ref{S_I}). 
Explicit calculation of the diagram shown in Fig. \ref{fig-Idqd-diag}d) leads to the final expression for DQD current  (\ref{I_DQD-Fano}). 

The suppression of the DQD current for $V_{\mathrm{QPC}}$ exceeding the
energy conservation threshold $\Delta/e$ represents a profound feature of the
current voltage characteristics of the
double dot system under consideration, which has no analogy in the 
shot noise induced current through a single quantum dot \cite{Onac06}. 
It can  be understood in simple terms, if one considers the generation of DQD
current as a kind of a Coulomb drag experiment. Indeed, the diagrammatic representation
for the inelastic current Fig. \ref{fig-Idqd-diag}d) is almost identical to the
diagramms for the drag current,  the difference being the presence of a
nonequilibrium polarization operator $\Pi$ for QPC instead of the current
operator \cite{Kamenev-Levchenko,drag}. To realize the drag, the particle-hole (p-h) symmetry has to be
violated in both components of the system. Its violation in the gated DQD is explicit. In the case of symmetric circuit, the violation of the p-h symmetry in
the nonequilibrium QPC is possible only if the transmission amplitudes at
energies differing by $\Delta$ are different. Indeed, the initial bosonic
excitation that provides the energy to the DQD is an electron-hole pair in QPC
with energy $\Delta$. The Hamiltonian describing the propagation
of such a pair through QPC is p-h asymmetric only for different transmission
amplitudes for the electrons and the holes.

\begin{acknowledgments} 
The author is grateful to L. Glazman, A. Kamenev, and A. Levchenko 
for illuminating discussions. 
The author appreciates and enjoyed the hospitality  of the William I. Fine
Theoretical Physics Institute, University of
Minnesota, where a part of this work has been performed. 
The author acknowledges Financial support from DFG through
Sonderforschungsbereich 508 and Sonderforschungsbereich 668. 
\end{acknowledgments}


\begin{thebibliography}{99}
\bibitem{indices} Note that index R/L relates to the reservoir. It should not be mixed with chirality that is not conserved in the presence of potential scattering.
\end{thebibliography}
\end{document}